\documentclass[aps, prl, twocolumn, groupedaddress, showpacs, floatfix,10pt]{revtex4-1}

\usepackage{amsmath, amssymb}
\usepackage[english]{babel}
\usepackage[latin1]{inputenc}
\usepackage{xy}
\usepackage{graphicx}
\usepackage{verbatim}
\usepackage{enumerate}
\usepackage{url}
\usepackage{subfigure}
\usepackage{bm}
\usepackage{times}

\newcommand{\imagescaling}{0.8}

\newcommand{\note}[1]{}

\newcommand{\twocol}[1]{#1}

\newcommand{\maxdim}{N}
\newcommand{\maxpop}{M}

\newcommand{\Splay}{\Dp}
\newcommand{\Sync}{\Sp}

\newcommand{\vth}{\vartheta}

\newcommand{\e}{\varepsilon}
\newcommand{\eps}{\e}

\newcommand{\R}{\mathbb{R}}

\newcommand{\Z}{\mathbb{Z}}

\newcommand{\Tor}{\mathbf{T}}

\newcommand{\Tormn}{\Tor^{\maxpop\maxdim}}

\newcommand{\sset}[1]{\left\lbrace #1\right\rbrace}

\newcommand{\Sp}{\textrm{S}}
\newcommand{\Dp}{\textrm{D}}

\newcommand{\SSp}{\Sync\Sync}
\newcommand{\BSS}{\mathcal{B}(\SSp)}

\newcommand{\TS}{\Tor\Sync}

\newcommand{\omBr}{\omega_{\text{Br}}}

\newcommand{\taBr}{\tau_{\text{Br}}}
\newcommand{\taEx}{\tau_{\text{Ex}}}
\newcommand{\gBr}{g_{\text{Br}}}
\newcommand{\gEx}{g_{\text{Ex}}}

\newcommand{\gbEx}{\bar g_{\text{Ex}}}

\begin{document}

\preprint{10.1103/PhysRevLett.119.168301}


\title{
Robust Weak Chimeras in Oscillator Networks with Delayed Linear and Quadratic Interactions
}
\author{Christian Bick${}^{a,b}$, Michael Sebek${}^{c}$, Istv\'an Z.~Kiss${}^{c}$}%

\affiliation{%
\mbox
{${}^a$Oxford Centre for Industrial and Applied Mathematics, Mathematical Institute, University of Oxford, Oxford OX2 6GG, UK}\\
\mbox
{${}^b$Centre for Systems Dynamics and Control and Department of Mathematics, University of Exeter, Exeter EX4~4QF, UK}\\
${}^c$Department of Chemistry, Saint Louis University, 3501 Laclede Ave., St.~Louis, MO 63103, USA}
\date{\today}


\begin{abstract}
We present an approach to generate chimera dynamics (localized frequency synchrony) in oscillator networks with two populations of (at least) two elements using a general method based on delayed interactions with linear and quadratic terms. The coupling design yields robust chimeras through a phase-model-based design of the delay and the ratio of linear and quadratic components of the interactions. We demonstrate the method in the Brusselator model and experiments with electrochemical oscillators. The technique opens the way to directly bridge chimera dynamics in phase models and real-world oscillator networks.
\end{abstract}

\maketitle

\noindent
Phase models provide mathematical descriptions of weakly coupled oscillatory systems. The state of each unit is described by a single variable, its phase, and the effect of coupling is determined by the phase velocity as a function of the phase difference of the coupled elements~\cite{Kuramoto, Acebron2005, Ashwin1992}. They capture collective dynamical phenomena (e.g., synchronization and dynamical differentiation) of even very large networks of oscillators, as it was demonstrated with electrochemical~\cite{Kiss:2005PM} and neural oscillations~\cite{Hansel1993a, Ashwin2015}, and superconducting Josephson junctions~\cite{Wiesenfeld1998}. Phase-model-based approaches have also been effective to induce desirable synchronization patterns with external signals, e.g., with desynchronization, and stable and itinerant cluster dynamics~\cite{Kiss2007, Pikovsky2004_desync, Orosz:2009fr}.

Many biological systems, however, operate at an intermediate level of (frequency) synchronization~\cite{Uhlhaas:2006ys}. Collective dynamics where oscillators are only locally frequency synchronized---commonly know as chimeras---are striking examples for the rich dynamics that arise even in identical units~\cite{Panaggio2015, Scholl2016} that are of relevance in applications~\cite{Bick2014a}. Much theoretical effort has focused on chimeras in phase oscillator networks. These range from explicit bifurcation analyses~\cite{Abrams2008} to a mathematically rigorous notion of a chimera---a weak chimera is characterized by angular frequency synchronization along trajectories---and corresponding existence results~\cite{Ashwin2014a, Bick2015c, Bick2015d}. At the same time, carefully designed experiments with chimera dynamics have only drawn inspiration from the phase oscillator results~\cite{Tinsley2012, Hagerstrom2012, Martens2013, Wickramasinghe:2013zr} rather than relate directly to them. Indeed, general experimental realization conditions for robust chimeras (as asymptotic dynamics which arise despite the inherent heterogeneities) are difficult to formulate because of the complexities of the experimental systems. For example, the electrochemical chimera system \cite{Wickramasinghe:2013zr} lasted only 100 cycles, required many connections (at least 20 units with 140 connections) and showed chimera dynamics with unrealistically uniform system with natural frequency differences less than $0.1\%$. 

In this Letter, we show that very robust chimeras arise in a small oscillator network of only two populations of two elements, when the interactions among the elements are designed in general way with weak linear and quadratic, time-delayed interactions. The interactions are based on a phase model for which we predict the emergence and bifurcations of weak chimeras. The effective design is achieved by the generalization of a feedback approach previously used to induce collective dynamics of globally coupled networks~\cite{Kiss2007, Kori2008, Rusin2010} to complex network structures. We verify our approach in numerical simulations of the Brusselator model and experiments with electrochemical oscillators to observe weak chimeras in these systems.


\textit{Weak chimeras in networks of phase oscillators---}%
We consider the dynamics of $\maxpop=2$ populations of $\maxdim=2$ phase oscillators where the phase interaction between oscillators is determined by the coupling function
\begin{equation}\label{eq:CplngBistab}
g(\phi) = \sin(\phi-\alpha) + r\sin(2(\phi-\alpha))
\end{equation}
with parameters $\alpha, r\in\R$. More precisely, let the phase~$\theta_{\sigma, k}\in \Tor := \R/2\pi\Z$ of oscillator~$k$ in population~$\sigma\in\sset{1,2}$ evolve according to
{\allowdisplaybreaks
\begin{subequations}\label{eq:OscPopulations}
\begin{align}
\begin{split}
\dot\theta_{\sigma, 1} &= \omega + g(\theta_{\sigma, 2}-\theta_{\sigma, 1}) 
\twocol{\\&\qquad\qquad}
+\eps \left(g(\theta_{\kappa, 1}-\theta_{\sigma, 1})+g(\theta_{\kappa, 2}-\theta_{\sigma, 1})\right)
\end{split}\\
\begin{split}
\dot\theta_{\sigma, 2} &= \omega + g(\theta_{\sigma, 1}-\theta_{\sigma, 2}) 
\twocol{\\&\qquad\qquad}
+\eps \left(g(\theta_{\kappa, 1}-\theta_{\sigma, 2})+g(\theta_{\kappa, 2}-\theta_{\sigma, 2})\right)
\end{split}
\end{align}
\end{subequations}
}%
where $\kappa = 3-\sigma$, $\omega=1$ is the intrinsic frequency of each oscillator%
~\footnote{Note that we can set $\omega=1$ without loss of generality in the phase equations~\eqref{eq:OscPopulations} by going into a suitable co-rotating frame. Moreover, in the phase dynamics we omit the global coupling strength~$K$ introduced later since we can assume $K=1$ by rescaling time appropriately.}, and $\eps$ is the interpopulation coupling parameter;
see Fig.~1(a) of the Supplemental Material for a sketch of the network topology. If~$\theta(t)$ is a trajectory of~\eqref{eq:OscPopulations} with initial condition $\theta(0)=\theta^0$ then let $\hat\theta(t)$ be a continuous lift of~$\theta$ to~$\R$. With $\Omega_{\sigma, k}(T):= \frac{1}{T}\hat\theta_{\sigma, k}(T)$ we have the asymptotic average angular frequency $\Omega_{\sigma, k}=\lim_{T\to\infty}\Omega_{\sigma, k}(T)$ of oscillator $(\sigma, k)$. Recall that the characterizing feature of a \emph{weak chimera} as a particular invariant set~$A\subset\Tormn$ is \emph{frequency synchrony} (and lack thereof): For all trajectories with initial conditions $\theta^0\in A$ we have distinct oscillators $(\sigma,k), (\eta,j), (\rho,\ell)$ such that $\Omega_{\sigma,k}=\Omega_{\eta,j}\neq\Omega_{\rho,\ell}$; see~\cite{Ashwin2014a, Bick2015c, Bick2015d} for a precise definition.

If $\eps=0$, the populations in~\eqref{eq:OscPopulations} are uncoupled, which gives rise to invariant subspaces. Each population evolves on~$\Tor^2$, and, for a moment, we suppress the population index~$\sigma$. The set $\Sync = \sset{\theta_1 = \theta_2}$ corresponds to full phase synchrony and $\Splay = \sset{\theta_1 = \theta_2+\pi}$ denotes the splay phase where oscillators are in antiphase. The asymptotic average frequencies of the oscillators can be written in terms of the coupling function~$g$: We have $\Omega_{k}(\theta^0) = \omega + g(0)$ for $\theta^0\in\Sync$ and $\Omega_{k}(\theta^0) = \omega+g(\pi)$ for $\theta^0\in\Splay$. Moreover, $g$ determines the stability of~$\Sync$ and~$\Splay$. If~$g$ has only a single harmonic, $r=0$, then full synchrony~$\Sync$ and~$\Splay$ exchange stability at $\alpha = \pm\frac{\pi}{2}$ in a degenerate bifurcation. A second nontrivial harmonic, $r\neq 0$, breaks this degeneracy, that is, for~$\alpha\approx\pm\frac{\pi}{2}$ there is a branch of stable (relative) equilibria for $r>0$~\cite{Rusin2010} and a region of bistability between~$\Sync$ and~$\Splay$ for $r<0$.

\begin{figure}
\subfigure[\ Phase plane for $\e=0.1$]{
\includegraphics[scale=\imagescaling]{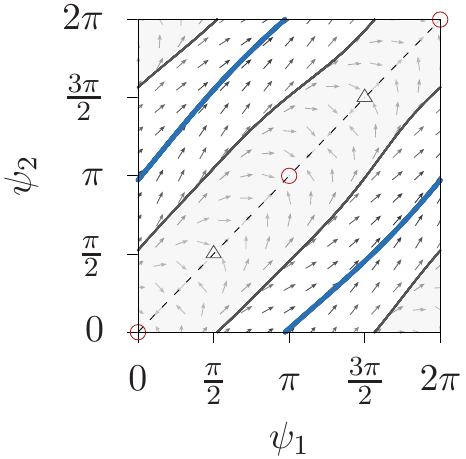}}\hfill
\subfigure[\ Frequency synchrony for varying~$\e$]{
\includegraphics[scale=\imagescaling]{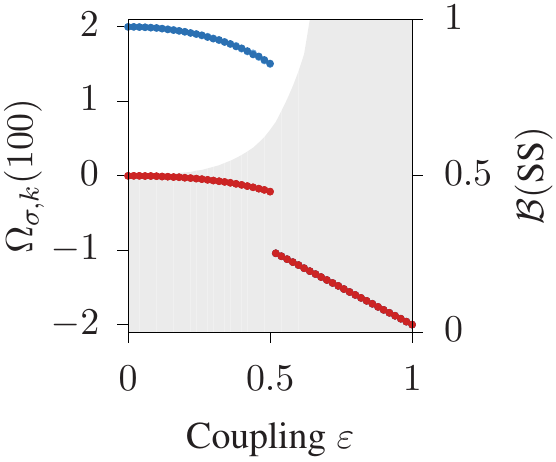}}
\caption{\label{fig:PhaseOscillators}
Weak chimeras exist for two coupled populations~\eqref{eq:OscPopulations} of $\maxdim=2$ oscillators. Panel~(a) shows the phase plane of the reduced system~\eqref{eq:OscPopulationsRestr} for $\e=0.1$ (shading of arrows indicates their norm): Stable phase synchronized solutions in $\SSp$ are shown in red, the stable weak chimera periodic orbit in blue. Unstable periodic orbits (gray) bound both~$\BSS$ (shaded area) and the basin of attraction of the weak chimera.  In Panel~(b), the shading indicates the fraction of initial conditions $(\psi_1(0), \psi_2(0))$ that lie in~$\BSS$. Separation of average angular frequencies ($\Omega_{1,k}$ in blue and $\Omega_{2,k}$ in red) characterizes the weak chimeras for~\eqref{eq:OscPopulations} with fixed initial condition $(\theta_{1,1}(0), \theta_{1,2}(0), \theta_{2,1}(0), \theta_{2,2}(0)) = \left(0, \pi, \frac{\pi}{2}, \frac{\pi}{2}+0.1\right)$. Note that, while the stable weak chimera exists up to $\e\approx 0.64$, this initial condition leaves its basin of attraction at $\e\approx 0.5$.}
\end{figure}

For phase shifts~$\alpha\approx\frac{\pi}{2}$ and $r<0$, the system~\eqref{eq:OscPopulations} now supports weak chimeras for a wide range of parameter values $\eps>0$. Such chimeras arise as perturbations of $\Splay\times\Sync$~\footnote{Since the system is symmetric with respect to permuting the populations there are also weak chimeras close to $\Sync\times\Splay$ for small $\eps$.} for small $\eps>0$~\cite{Ashwin2014a}. For the dynamics~\eqref{eq:OscPopulations} the space $\TS:=\Tor^2\times\Sync = \sset{\theta_{2, 1}=\theta_{2, 2} =: \vth}$, where the second population is phasesynchronized, is dynamically invariant and its transversal stability is determined by~$g'(0)$~\cite{Ashwin1992, Bick2015c}. The dynamics on~$\TS$ are determined by the phase differences $\psi_k := \theta_{1, k}-\vth$ which evolve as
\begin{subequations}\label{eq:OscPopulationsRestr}
\begin{align}
\begin{split}
\dot\psi_1 &= g(\psi_2-\psi_1) - g(0)
\twocol{\\&\qquad}
+\eps\left( 2g(-\psi_1) - g(\psi_1) - g(\psi_2)\right)
\end{split}\\
\begin{split}
\dot\psi_2 &= g(\psi_1-\psi_2) - g(0)
\twocol{\\&\qquad}
+\eps \left(2g(-\psi_2) - g(\psi_1) - g(\psi_2)\right).
\end{split}
\end{align}
\end{subequations}
in a frame that rotates (not necessarily uniformly) with the synchronized population~$\theta_{\sigma, k}$. The set $\SSp := \Sync\times\Sync = \sset{\psi_1=\psi_2}\subset \TS$ is dynamically invariant.
Fig.~\ref{fig:PhaseOscillators}(a) shows the stable limit cycle that corresponds to the stable weak chimera in the full system and in-phase and antiphase synchronized clusters in the phase plane of~\eqref{eq:OscPopulationsRestr} for $\alpha=1.57$, $r=-0.3$, and $\e = 0.1$. For initial conditions in~$\TS$, trajectories converge either to the weak chimera or to equilibria on~$\SSp$; thus, we calculated the size of the basin of attraction of the weak chimera periodic orbit as the complement of the basin of attraction~$\BSS$~\footnote{We estimated the size of~$\BSS$ numerically by checking the fraction of whether a trajectory with initial condition on a uniform grid had approached~$\SSp$ after $T=300$ time units up to tolerance $10^{-3}$.}. The basin of attraction of the stable periodic orbit shrinks as $\e \geq 0$ is increased---shown in Fig.~\ref{fig:PhaseOscillators}(b)---and the periodic orbit becomes unstable in a pitchfork bifurcation of limit cycles at $\e\approx 0.64$. The unstable limit cycle is ultimately destroyed at $\e\approx 0.715$ in a global bifurcation. The stable weak chimeras here are robust against small perturbations of the system (as hyperbolic limit cycles) and, in contrast to systems where~$g$ has only a single harmonic~\cite{Panaggio2015b}, exist for a wide range of parameters with a relatively large basin of attraction. This makes them suitable for realization in limit cycle oscillators through feedback.

\textit{Feedback induces weak chimeras in the Brusselator---}%
We first illustrate our engineering approach to obtain weak chimeras in limit cycle oscillator systems using the Brusselator model, a simple two-variable ordinary differential equation system that admits a Hopf bifurcation~\cite{Glansdorff1971}. For real parameters $A$ and $B$, define $f(x, y) = \frac{B}{A}x^2 + 2Axy + x^2y$. Let $p_{\sigma, k}(t)$ be a control signal for the $k$th oscillator in population $\sigma\in\sset{1, 2}$ whose dynamics are given by
\begin{subequations}\label{eq:Brusselator}
\begin{align}
\dot x_{\sigma, k} &= (B-1)x_{\sigma, k} + A^2y_{\sigma, k} + f(x_{\sigma, k}, y_{\sigma, k}) + K p_{\sigma, k}(t)\\
\dot y_{\sigma, k} &= -Bx_{\sigma, k} - A^2y_{\sigma, k} - f(x_{\sigma, k}, y_{\sigma, k}).
\end{align}
\end{subequations}
where~$K$ is the total gain for the control signal that is applied to the first component. Fix $A = 1$ and $B = 2.3$. For $K=0$, each oscillator has a stable limit cycle with angular speed $\omBr=0.977$ and period $T_{\text{Br}} = 2\tau_{\text{Br}} = 2\pi\omBr^{-1}$.

For appropriately chosen control and sufficiently small~$K$, the network of Brusselator oscillators has a desired phase reduction; see also~\cite{Kori2008}. More precisely, given a uniformly increasing phase variable $\dot\phi_{\sigma, k} = 1$ on the limit cycle for $K=0$ and a target interaction function $g(\phi) = \sum_{\ell\in\Z} g_\ell\exp(-i\ell\phi)$ for the phase dynamics, then the feedback $h(\phi) = \sum_{\ell\in\Z} h_\ell\exp(-i\ell\phi)$ can be obtained from the phase response curve $Z(\phi) = \sum_{\ell\in\Z} Z_\ell\exp(-i\ell\phi)$ by solving 
\begin{equation}\label{eq:CplngFuncFourier}
g_\ell = Z_{-\ell}h_\ell.
\end{equation}
For feedback control, the~$h_\ell$ may be expressed in terms of  the waveform $x(\phi) = \sum_{\ell\in\Z} a_\ell\exp(-i\ell\phi)$ which yields a set of equations involving the Fourier coefficients of the waveform, the phase response curve, and the target interaction function~$g$.

\begin{figure*}
\subfigure[\ Interaction function]{\includegraphics[scale=\imagescaling]{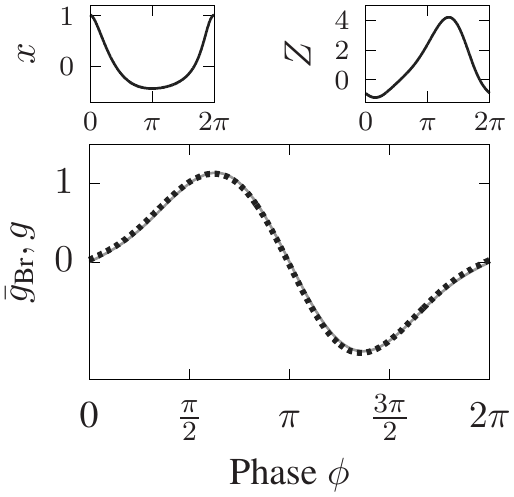}}
\qquad
\subfigure[\ Brusselator weak chimera]{\includegraphics[scale=\imagescaling]{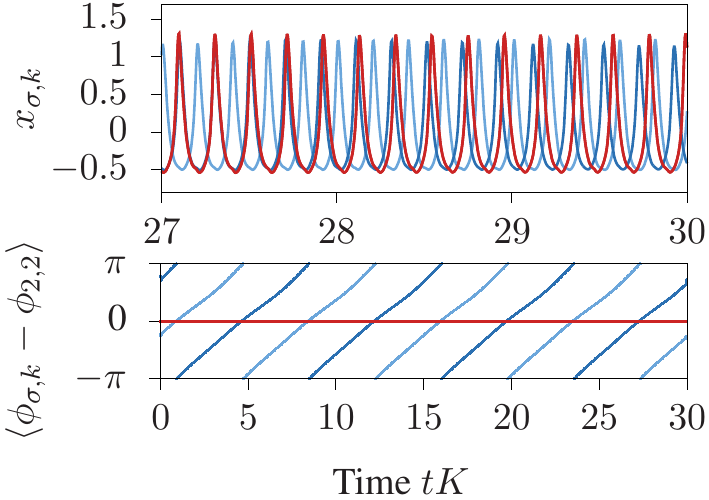}}
\qquad
\subfigure[\ Brusselator bifurcation diagram]{
\includegraphics[scale=\imagescaling]{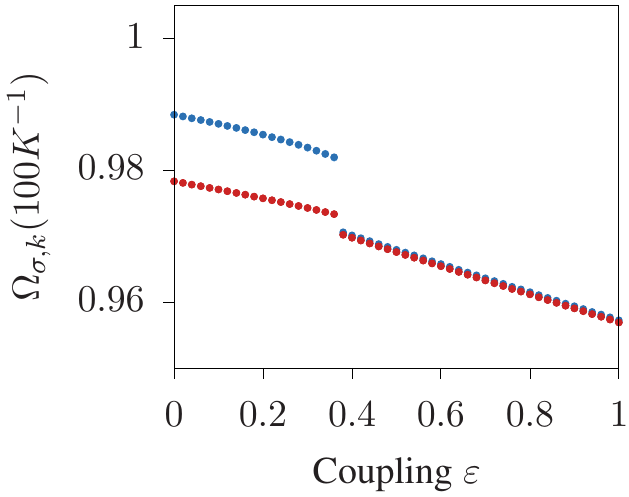}}
\caption{\label{fig:Brusselator}
Network interaction yields weak chimeras in Brusselator models for $A=1$ and $B=2.3$. 
Panel~(a) shows the waveform~$x$, phase response curve~$Z$, and the effective interaction function~$\gBr$ (dotted) on top of the target~\eqref{eq:CplngBistab} with $\alpha=0$ (solid) for second-order feedback parameters $(\tau_1, \tau_2) = 2\pi\omBr^{-1}(0.8577, 0.3156)$ and $(k_1, k_2)= (0.6601, -2.1692)$. 
Panel~(b) depicts waveforms $x_{\sigma, k}$ and phase differences $\langle\phi_{k, \sigma}-\phi_{2,2}\rangle$ of a weak chimera for this feedback, $\eps=0.1$, and $K=0.03$: The frequencies of populations~1 (blue) and 2 (red) are distinct.
These weak chimeras exist for a range of~$\eps$ as shown in Panel~(c), here $K=0.05$. The initial condition was the point with $(\phi_{1,1}(0), \phi_{1,2}(0), \phi_{2,1}(0), \phi_{2,2}(0)) = \left(0, \pi, \frac{\pi}{2}, \frac{\pi}{2}+0.1\right)$ for the uncoupled system.}
\end{figure*}

We realized the network~\eqref{eq:OscPopulations}---up to a rescaling of time---with the feedback signal
\begin{equation}
p_{\sigma, k}(t) = \sum_{\kappa,j\in\sset{1,2}} K_{\kappa\sigma}h(x_{\kappa, j}(t-\tau)),
\end{equation}
where the feedback gains $K_{\kappa\sigma} = 1$ if $\kappa=\sigma$ and $K_{\kappa\sigma} = \e$ otherwise determine the network topology and~$\tau$ is a global feedback delay. Let $\bar x_{\kappa,j} = x_{\kappa,j}-a_0$, where~$a_0$ is the zeroth Fourier coefficient of~$x_{\kappa,j}$ as above (similarly, write $\bar g=g-g_0$). Now for a given coupling function~\eqref{eq:CplngBistab} and second order time-delayed feedback
\begin{equation}\label{eq:FeedbackBrus}
\begin{split}
h(x_{\kappa,j}(t)) &= k_1\!\left(\bar x_{\kappa,j}(t-\tau_1) - \bar x_{\kappa,j}(t-\tau_1-\tau_{\text{Br}})\right)
\\&\qquad
+k_2\!\left((\bar x_{\kappa,j}(t-\tau_2)^2 + \bar x_{\kappa,j}(t-\tau_2-\tau_{\text{Br}})^2\right),
\end{split}
\end{equation}
the feedback parameters~$k_s$ and~$\tau_s$ are readily computed~\cite{Kiss2007}. Since the delay~$\taBr$ is equal to half of the oscillations period, the linear feedback term will not have a second harmonic and the quadratic term will not have a first harmonic. Moreover, each delay effectively acts as a phase shift for the coupling function (as long as $K\tau$ is small)~\cite{Kiss2007, Kori2008}. Thus, choosing~$\tau_1$ and~$\tau_2$ to yield pure~$\sin$ interaction for the first- and second-order feedback and then setting $k_2/k_1=r$ yields the coupling function~\eqref{eq:CplngBistab} for $\alpha=0$. For $\alpha\neq 0$, set the global delay to $\tau=\omBr^{-1}\alpha$. This strategy yields a good approximation~$\gBr$ of the target interaction function~\eqref{eq:CplngBistab} as shown in Fig.~\ref{fig:Brusselator}(a)~\footnote{The phase response curve was calculated using XPP~\cite{Ermentrout2002} to evaluate~\eqref{eq:CplngFuncFourier}.} without the need for extensive nonlinear fitting as in previous approaches~\cite{Kiss2007}.

Subject to feedback, the network of Brusselator oscillators~\eqref{eq:Brusselator} gives rise to weak chimeras. Fig.~\ref{fig:Brusselator}(b) shows the evolution of~$x_{\sigma, k}$ and~$\langle\phi_{\sigma, k}\rangle$---the average of $\phi_{\sigma, k}$ over one cycle---for $\e=0.1$. The average angular frequencies $\Omega_{\sigma, k}$ (calculated from the phase~$\phi_{\sigma, k}$ of each oscillator) of the synchronized and antiphase population are distinct. Stable weak chimeras exist for a range of coupling parameters~$\e$ as shown in Fig.~\ref{fig:Brusselator}(c), comparable to the predictions obtained from the phase oscillator dynamics~\eqref{eq:OscPopulations}. The weak chimeras are robust to adding small variations in the oscillators (not shown).


\textit{Experimental realization in electrochemical oscillators---}%
We built an experimental system with four (two populations of two) oscillatory chemical reactions that can be coupled through linear and quadratic feedback with a delay. Each oscillatory reaction occurs on the surface of a $1.00~\mathrm{mm}$ diameter nickel wire in $3~\mathrm{M}$ sulfuric acid. Because the disk electrodes are placed far from each other (about $3~\mathrm{mm}$ spacing) and the potential drop in the electrolyte is very small (about $0.1~\mathrm{mV}$), the oscillators do not show synchronization without the presence of additional coupling means~\cite{Kiss2002}. A multichannel potentiostat interfaced with a real-time Labview controller sets the potential $V_{\sigma, k}(t)$ of the wires individually with respect to a $\textrm{Hg/Hg}_2\textrm{SO}_4/\textrm{sat.~K}_2\textrm{SO}_4$ reference electrode. The currents $I_{\sigma, k}(t)$ of the four electrodes and a Pt-coated Ti electrode are recorded and converted to electrode potentials $E_{\sigma, k}(t) = V_{\sigma, k}(t) - I_{\sigma, k}(t)R_{\textrm{ind}}$, where $R_{\textrm{ind}} = 1~\mathrm{kOhm}$ is an individual resistance attached to each wire~\cite{Kiss2007}; see Supplementary Material for more details on the experimental setup. The electrode potential is corrected for offset, $\bar{E}_{\sigma, k} = E_{\sigma, k} - o$, where~$o$ is a time averaged electrode potential. For the feedback, the circuit potentials of each wire is set to
\begin{equation}\label{couple}
V_{\sigma, k}(t) = V_ {0} + K \sum_{\kappa,j\in\sset{1,2}} K_{\kappa\sigma}h(\bar{E}_{\kappa,j}(t-\tau)),
\end{equation}
where $K_{\kappa\sigma}$ determines the network topology, $K$ is the total feedback gain, $\tau$ is the global delay, and
\begin{equation}\label{eq:FeedbackExp}
\begin{split}
h(E_{\kappa,j}(t)) &= k_1\!\left(\bar E_{\kappa,j}(t-\tau_1) - \bar E_{\kappa,j}(t-\tau_1-\taEx)\right)
\\&\qquad
+k_2\!\left((\bar E_{\kappa,j}(t-\tau_2)^2 + \bar E_{\kappa,j}(t-\tau_2-\taEx)^2\right)
\end{split}
\end{equation}
is the feedback as in~\eqref{eq:FeedbackBrus}. As above, the delay~$\taEx$ is set to be equal to half a period of the uncoupled oscillators: For the experimental setup with no coupling, $K=0$, and potential set to $V_{0} = 1160~\mathrm{mV}$ the electrodissolution process is oscillatory with a natural frequency of about $0.45~\mathrm{Hz}$. During the experiment, a 2---3$~\mathrm{mHz}$ difference in natural frequencies between the electrodes was maintained.

\begin{figure}
\begin{center}
\includegraphics[width=1\columnwidth]{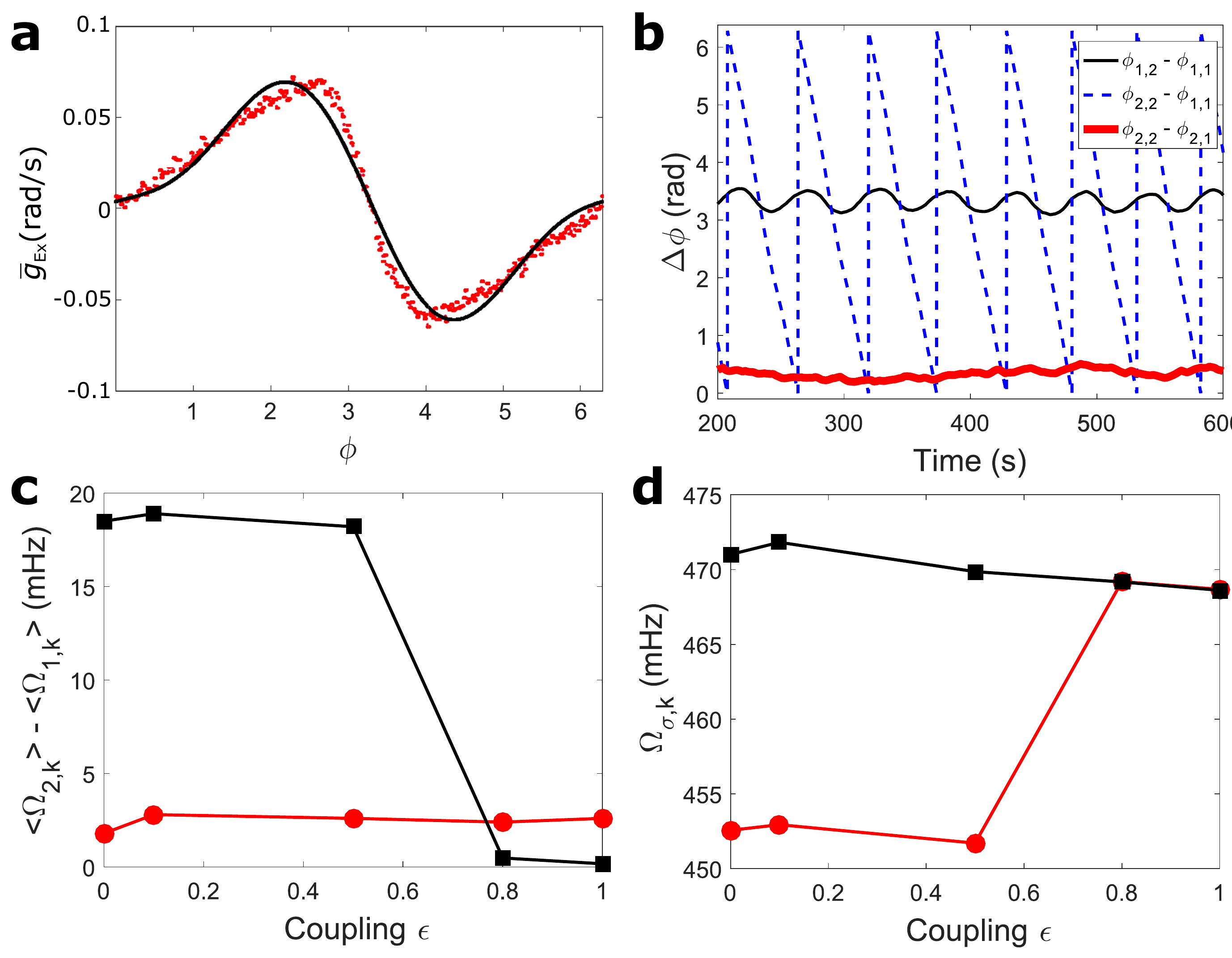}
\end{center}
\caption{Experimental weak chimera; $V_{0} = 1160~\mathrm{mV}$, $R_{\mathrm{ind}} \simeq 1~\mathrm{k}\Omega$.   
{(a)} Experimentally determined (points) and desired (line) interaction function for $\tau = 0$, $r=-0.4$; $K = 0.35$.
{(b)} Phase difference time series of population~1 ($\phi_{1,2} - \phi_{1,1}$), population~2 ($\phi_{2,2} - \phi_{2, 1}$), and between the populations ($\phi_{2,2} - \phi_{1,1}$) of a weak chimera; $K = 0.52$, $\eps = 0.1$.
{(c)} Differences of frequencies (averaged over populations) between populations without coupling, $K = 0$ (circles) and at various $\eps$ values, $K = 0.52$ (squares).
{(d)} The frequencies of the populations at various~$\eps$ values (squares = population~1, circles = population~2) at $K = 0.52$.}
\label{figure1}
\end{figure}

Initial trials allowed us to determine the feedback parameters to get the desired coupling functions~\eqref{eq:CplngBistab} using the same strategy as in the numerical simulations. Employing pure first- and second-order feedback gains, we set $k_{1} = 0.22$, $k_{2} = 2.0~\mathrm{V}^{-1}$, $\tau_1=\tau_2=\tau = 0$. With these parameters, we determined the phase interaction function (using a self-feedback method~\cite{Rusin2010}). Fig.~\ref{figure1}(a) shows that the experimental phase interaction function~$\gEx$ approximates the desired interaction function~\eqref{eq:CplngBistab} with $r=-0.4$ and $\alpha=0$ very well. In terms of Fourier coefficients, we obtained 
\begin{equation}\label{interaction}
\begin{split}
\gbEx(\phi) &= -0.012\cos(\phi) + 0.051 \sin(\phi) 
\twocol{\\&\qquad}
+0.003 \cos(2\phi) - 0.021\sin(2\phi)
\end{split}
\end{equation}
which shows that there are weak cosine and strong first and second harmonic sinusoidal components with $r=-0.41$. Adding a global delay of $\tau = 0.51~\mathrm{s}$, the uncoupled populations, $\eps=0$, exhibited bistability between in-phase and antiphase oscillatory states for $K=0.52$ (see Supplementary Material). This choice of parameters corresponds to a phase shift of $\alpha = 1.44$ in the phase model, is expected to fall within the chimera regime and was used in all the following experiments. Before the experiments, the phases of oscillators in populations~1 and~2 were set to anti- and in-phase configurations, respectively, to provide appropriate initial conditions for the weak chimera.


We observed weak chimeras in the experimental setup for a range of coupling parameters $\eps\geq 0$. First note that, if there are no interpopulation connections, $\eps = 0$, there is a very large dynamically induced frequency difference of about $18~\mathrm{mHz}$ between population~1 in an antiphase and population~2 in an in-phase configuration; see Figs.~\ref{figure1}(c) and 3(d). When the coupling between the populations was increased to $\eps = 0.1$, the populations remain approximately in the anti- and in-phase configurations [see Fig.~\ref{figure1}(b)] but now exhibited oscillations due to the interaction between populations. Importantly, the two populations exhibited phase drifting behavior relative to each other; this state thus represents a weak chimera state. As is shown in Fig.~\ref{figure1}(c), the frequency difference between the populations in the chimera state is much larger (about 9~times) than the frequency difference without interpopulation coupling. We observed a chimera state for a large interpopulation coupling strength up to $\eps = 0.5$; see Figs.~\ref{figure1}(c) and 3(d). As~$\eps$ was increased, the amplitude of the phase difference oscillations of the synchronized population increased. With strong interpopulation coupling at $\eps = 0.8$ the weak chimera breaks down and the two populations became phase locked (see Supplementary Material).


\textit{Discussion---}%
We showed that a simple network of two populations of two elements, coupled through a linear and quadratic amplification with a delay of half the period, can generate very robust chimera patterns with strong phase slipping behavior between the populations. The induced chimeras do not rely on amplitude dynamics (e.g., from chaotic~\cite{Hart2016} or amplitude~\cite{Schmidt:2014ly} clusters). Similar dynamics are expected with any nonlinear oscillatory system with a phase interaction function that has strong first harmonic and weak second harmonic components. While oscillations very close to a Hopf bifurcation typically have dominant first harmonics in the interaction functions, second harmonics arise naturally away from the Hopf bifurcation point (unless there is some particular symmetry in the individual dynamics)~\cite{Kori:2014mz}. Therefore, we expect that systems in which the oscillations occur through Hopf bifurcations (e.g., class 2 neurons or resonators \cite{Izhikevich:2000vn}) can exhibit the chimeras. Moreover, interactions between oscillators are often nonlinear in nature; examples include gating mechanisms that limit the transduction of the coupling signals  signals, as was demonstrated with glycolytic oscillations~\cite{Dhumpa:2014fs} or through synaptic coupling of neurons~\cite{neurobook}. Thus, our results give insights into how chimera dynamics emerge from the first two orders of the interaction, or could be induced if these interactions can be tuned.

Our results directly bridge complex collective dynamics, such as chimeras, in phase models and real-world oscillator networks. On the one hand, this approach elucidates how complex collective dynamics, which arise in phase oscillator networks, translate into networks of limit cycle oscillators. On the other hand, we anticipate insights into the limitations of phase oscillator models to describe collective dynamics of real-world networks beyond clustering; for example, this should further clarify the need to include higher-order interaction terms~\cite{Rosenblum2007, Ashwin2015a, Bick2016b} in the phase dynamics as they facilitate spatiotemporal dynamics of localized synchrony~\cite{Bick2017c}.

\textit{Acknowledgements---}%
C.B. acknowledges the warm hospitality at SLU and helpful discussions with Peter Ashwin. C.B. has received funding from the People Programme (Marie Curie Actions) of the European Union's Seventh Framework Programme (FP7/2007--2013) under REA Grant Agreement No.~626111. I.Z.K. acknowledges support from National Science Foundation Grant No.~CHE-1465013.


%

\end{document}